\documentclass[useAMS, usenatbib, onecolumn]{mn2e}
\usepackage{epsfig}
\usepackage{amsmath}
\def\be{\begin{equation}}
\def\ee{\end{equation}}
\def\ba{\begin{eqnarray}}
\def\ea{\end{eqnarray}}
\def\go{\mathrel{\raise.3ex\hbox{$>$}\mkern-14mu
             \lower0.6ex\hbox{$\sim$}}}
\def\lo{\mathrel{\raise.3ex\hbox{$<$}\mkern-14mu
             \lower0.6ex\hbox{$\sim$}}}

\def\bu{{\bf u}}

\def\d{\partial}

\def\tomega{\tilde\omega}

\begin{document}

\title[Super-Reflection in Fluid Discs]
{Super-Reflection in Fluid Discs: Corotation Amplifier, Corotation
Resonance, Rossby Waves, and Overstable Modes} 

\author[D. Tsang and D. Lai]{David Tsang$^{1,2}$\thanks{Email:
dtsang@astro.cornell.edu; dong@astro.cornell.edu} and Dong
Lai$^{1}$\footnotemark[1] \\ 
$^1$Center for Radiophysics and Space Research, Department of Astronomy,
Cornell University, Ithaca, NY 14853, USA \\
$^2$Department of Physics,
Cornell University, Ithaca, NY 14853, USA \\}

\volume{387}\pagerange{446--462} \pubyear{2008}
\label{firstpage}

\maketitle

\begin{abstract}
In differentially rotating discs with no self-gravity, 
density waves cannot propagate around
the corotation, where the wave pattern rotation speed equals the fluid
rotation rate. Waves incident upon the corotation barrier may be 
super-reflected (commonly referred to as corotation amplifier), but 
the reflection can be strongly affected by wave absorptions at the corotation
resonance/singularity. The sign of the absorption is related to the 
Rossby wave zone very near the corotation radius. 
We derive the explicit expressions for the 
complex reflection and transmission coefficients, taking into account 
wave absorption at the corotation resonance. 
We show that for generic discs, this absorption plays a much more
important role than wave transmission across the corotation barrier.
Depending on the sign of
the gradient of the vortensity of the disc, 
$\zeta=\kappa^2/(2\Omega\Sigma)$
(where $\Omega$ is the rotation rate, $\kappa$ is the epicyclic frequency,
and $\Sigma$ is the surface density), the corotation resonance can 
either enhance or diminish the super-reflectivity, and this 
can be understood in terms of the location of the Rossby wave zone 
relative to the corotation radius. 
Our results provide the explicit conditions 
(in terms of disc thickness, rotation
profile and vortensity gradient) for which super-reflection can be 
achieved. Global overstable disc modes may be possible for discs 
with super-reflection at the corotation barrier.
\end{abstract}


\section{Introduction}

Differentially rotating fluid discs, ubiquitous in astrophysics, 
are known to exhibit rich dynamics and possible instabilities 
(e.g. Papaloizou \& Lin 1995; Balbus \& Hawley 1998).
While local instabilities, such as Rayleigh's centrifugal instability 
(for discs with specific angular momentum decreasing outwards),
gravitational instability (for self-gravitational discs with too
large a surface density, or more precisely, Toomre $Q\lo 1$), 
and magnetorotational instability 
(for discs with a sub-thermal magnetic field), are well 
understood (at least in the linear regime), global effects and 
instabilities are more subtle, since they involve couplings and feedbacks 
of fluid at different locations (see Goldreich 1988 for an 
introduction/review). A well-known example is the corotation
amplifier (e.g. Mark 1976; Narayan, Goldreich \& Goodman 1987), which arises
from the interaction across the corotation between waves carrying opposite 
signs of angular momentum. Much stronger corotation amplifications
(WASER -- wave amplification by the stimulated emission of radiation, 
and SWING amplifiers) can be achieved for self-gravitating 
discs (e.g., Goldreich \& Lynden-Bell 1965; Julian \& Toomre 1966; 
Lin \& Lau 1975; see Shu 1992 for a review). 
Another well-known example is the Papaloizou-Pringle instability 
in finite accretion tori (confined between two free surfaces), 
in which coupling between waves inside the
corotation with those outside, combined with reflecting inner and outer 
boundaries, leads to violent overstable modes (Papaloizou \& Pringle
1984; Goldreich et al.~1986). Recent works on global 
disc instabilities include the Rossby wave instability (for discs with a 
strong enough density or vortensity bump; Lovelace et al.~1999;
Li et al.~2000) and the accretion-ejection instability (for magnetized
discs; Tagger \& Pellat 1999, Tagger \& Varniere 2006).

In this paper we are interested in 2D fluid discs without self-gravity
and magnetic field. For disturbances of the form $e^{im\phi-i\omega t}$,
where $m>0$ and $\omega$ is the wave (angular) frequency
(and thus $\omega_p=\omega/m$ is the pattern frequency), 
the well-known WKB dispersion relation for density waves 
takes the form (e.g., Shu 1992)
\be
(\omega-m\Omega)^2=\tomega^2=\kappa^2+k_r^2c^2,
\label{eq:disp}
\ee
where $\Omega$ is the disc rotation frequency, $\tomega=\omega-m\Omega$
is the Doppler-shifted wave frequency, $\kappa$ is the radial 
epicyclic frequency, $k_r$ is the radial wavenumber, and $c$ is 
the sound speed. Thus waves can propagate either inside the
inner Lindblad resonance radius $r_{\rm IL}$ (defined by $\tomega=-\kappa$)
or outside the outer Lindblad resonance radius $r_{\rm OL}$
(defined by $\tomega=\kappa$),
while the region around the corotation radius $r_c$ (set by $\tomega=0$)
between $r_{\rm IL}$ and $r_{\rm OL}$
is evanescent. Since the wave inside $r_{\rm IL}$ has pattern speed
$\omega_p$ smaller than the fluid rotation rate $\Omega$, it carries
negative wave action (or angular momentum), while the wave outside
$r_{\rm OL}$ carries positive wave action. As a result, a wave incident
from small radii toward the corotation barrier will be super-reflected, 
(with the reflected wave having a larger amplitude than the incident wave
amplitude) if it can excite a wave on the other side of the corotation barrier. 
If there exists a reflecting boundary at the inner disc radius
$r_{\rm in}$, then a global overstable mode partially trapped between
$r_{\rm in}$ and $r_{\rm IL}$ will result (see, e.g. Narayan et al.~1987 for
specific examples in the shearing sheet model, and Goodman \& Evans,1999 
and Shu et al.~ 2000 for global mode analysis of singular isothermal discs).

The simple dispersion relation (\ref{eq:disp}), however, does not capture
an important effect in the disc, i.e., corotation resonance or 
corotation singularity. 
Near corotation $|\tomega|\ll\kappa$, the WKB dispersion relation for
the wave is [see equation (\ref{eq:disper}) below]
\be
\tomega={2\Omega k_\phi\over k_r^2+k_\phi^2
+\kappa^2/c^2}\left({d\over dr}\ln{\kappa^2
\over 2\Omega\Sigma}\right)_c,
\label{eq:disp2}\ee
where $k_\phi=m/r$ and $\Sigma$ is the surface density,
and the subscript ``c'' implies that the quantity is evaluated
at $r=r_c$. The quantity 
\be
\zeta\equiv {\kappa^2\over 2\Omega\Sigma}=
{(\nabla\times {\bf u_0})\cdot {\hat z}\over\Sigma}
\label{eq:zeta}\ee
is the vortensity of 
the (unperturbed) flow (where ${\bf u_0}$ is the flow velocity). 
The dispersion relation (\ref{eq:disp2}) describes Rossby waves, analogous
to those studied in geophysics (e.g. Pedlosky 1987)
\footnote{A Rossby wave propagating in the Earth's atmosphere satisfies the
dispersion relation $\tomega=(2k_\phi/k^2R)(\partial\Omega_3/
\partial\theta)$,
where $k^2=k_\phi^2+k_\theta^2$,
$\Omega_3=\Omega\cos\theta$ is the projection of the rotation rate on 
the local surface normal vector and $\theta$ is the polar angle 
(co-latitude).}.
For $k_r^2\gg \kappa^2/c^2$ and $k_r^2\gg k_\phi^2$,
we see that Rossby waves can propagate either outside the 
rotation radius $r_c$ (when $d\zeta/dr>0$) or inside $r_c$ (when
$d\zeta/dr<0$). In either case, we have $k_r\rightarrow\infty$ as $r\rightarrow
r_c$. This infinite wavenumber signifies wave absorption
(cf. Lynden-Bell \& Kalnajs 1972 in stellar dynamical context; 
Goldreich \& Tremaine 1979 in the context of wave excitation in discs by
a external periodic force; see also Kato 2003, Li et. al. 2003, and Zhang \& Lai 2006 for wave 
absorption at the corotation in 3D discs). 
At corotation, the wave pattern angular speed $\omega/m$ matches $\Omega$,
and there can be efficient energy transfer between the wave and the background 
flow, analogous to Landau damping in plasma physics. 
Narayan et al.~(1987) treated this effect as
a perturbation of the shearing sheet model, and showed that the corotational
absorption can convert neutral modes in a finite shearing sheet
into growing or decaying modes. Papaloizou \& Pringle (1987) used a WKB
method to examine the effect of wave absorption at corotation on the 
nonaxisymmetric modes in an unbound (with the outer boundary extending to 
infinity) cylindrical torus.

In this paper, we derive explicit expressions  
for the complex reflection coefficient and transmission coefficient 
for waves incident upon the corotation barrier. We pay particular 
attention to the behavior of 
perturbations near the corotation resonance/singularity. 
Our general expressions include both the effects of corotation amplifier 
and wave absorption at corotation (which depends on 
$d\zeta/dr$). We show explicitly that depending on the sign of
$d\zeta/dr$, the corotation resonance/singularity can either enhance or 
diminish the super-reflectivity, and this can be understood in terms of 
the location of the Rossby wave zone relative to the corotation radius.

Our paper is organized as follows. After presenting the general
perturbation equations (section 2), we discuss the the wave
dispersion relation and propagation diagram, and derive
the local solutions for the wave equation around the Lindblad resonances
and corotation resonance (section 3). We then construct global WKB
solution for the wave equation, and derive the wave reflection, transmission
and corotational damping coefficients in section 4. An alternative
derivation of the wave damping coefficient is presented in section 5.
Readers not interested in technical details can skip Sections 2-5 and
concentrate on Section 6, where we illustrate our results and discuss their
physical interpretations. Section 6.1 contains a numerical calculation of the
wave reflectivity across corotation and discusses the limitation of the WKB
analysis. We discuss how global overstable
modes may arise when super-reflection at the corotation is present in 
section 7 and conclude in section 8.

\section{Perturbation Equations}

We consider a geometrically thin gas disc and adopt cylindrical coordinate
system $(r,\phi,z)$. The unperturbed disc has an integrated 
surface density $\Sigma(r)$ and velocity ${\bf u_0} = (0, r\Omega, 0)$.
The flow is assumed to be barotropic, so that the integrated pressure $P$
depends only on $\Sigma$. Self gravity of the disc is neglected.

The linear perturbation equations for the flow read
\ba
&&{\partial \over\partial t}\delta\bu+(\bu_0\cdot\nabla)\delta\bu
+(\delta\bu\cdot\nabla)\bu_0
=-\nabla\delta h, \label{eq:u}\\
&& {\partial\over\partial t}\delta\Sigma+\nabla\cdot(\Sigma\,\delta \bu
+\bu_0\,\delta\Sigma)=0, 
\label{eq:rho}\ea 
where $\delta\Sigma,~\delta \bu$ and $\delta h=\delta P/\Sigma$ 
are the (Eulerian) perturbations of surface density, velocity and enthalpy,
respectively. For barotropic flow, $\delta h$ and $\delta\Sigma$ are related by
\be
\delta h=c^2{\delta\Sigma\over\Sigma},
\ee
where $c$ is the sound speed, with $c^2=dP/d\Sigma$. 

We assume that the $\phi$ and $t$ dependence of the perturbation are of 
the form
\be
\delta\bu,~\delta\Sigma,~\delta h \propto e^{ i m \phi-i\omega t},
\ee
where $m$ is a positive integer, and $\omega$ is the wave (angular) frequency.
We presume $\omega>0$ so that the pattern (angular) speed of the perturbation 
$\omega_p=\omega/m$ is positive (in the same direction as the flow rotation).
Note that we usually assume $\omega$ is real, except 
in section 3.2 (dealing with the perturbation near corotation)
where we include a small imaginary part ($\omega=\omega_r+i\omega_i$, with
$\omega_i>0$) to represent slowly growing disturbances.
The perturbation equations (\ref{eq:u})-(\ref{eq:rho}) become
\ba
&& -i\tomega {\Sigma\over c^2}\delta h + \frac{1}{r} \frac{\d}{\d r} 
(\Sigma r \delta u_r) + \frac{im}{r}\Sigma \delta u_\phi = 0,\label{eq:sig}\\
&& -i\tomega \delta u_r - 2\Omega \delta u_\phi = -{\partial\over\partial r}
\delta h,\label{eq:ur}\\
&& -i\tomega \delta u_\phi + \frac{\kappa^2}{2\Omega}\delta u_r = 
-\frac{im}{r} \delta h,\label{eq:uphi}
\ea
where the epicyclic
frequency $\kappa$ is given by
\be
\kappa^2={2\Omega\over r}{d\over dr}(r^2\Omega).
\ee

Eliminating $\delta u_r$ and $\delta u_\phi$ from 
equations (\ref{eq:sig})-(\ref{eq:uphi}), we obtain a standard second-order 
differential equation governing $\delta h$ (e.g., 
Goldreich \& Tremaine 1979):
\be
\left[\frac{d^2}{dr^2} -
\frac{d}{dr}\left(\ln\frac{D}{r\Sigma}\right)\frac{d}{dr} -
\frac{2m\Omega}{r\tomega}\left(\frac{d}{dr}\ln\frac{\Omega\Sigma}{D}\right)
- \frac{m^2}{r^2} - \frac{D}{c^2}\right]\delta h = 0,\label{perturbeq1}
\ee
where 
\be
D \equiv \kappa^2 - \tomega^2=\kappa^2-(\omega-m\Omega)^2.
\ee
Defining 
\be
S = D/(r\Sigma),\quad 
\eta =S^{-1/2}\delta h, 
\ee
we can rewrite \eqref{perturbeq1} as a wave equation
\be
\left[\frac{d^2}{dr^2}-\frac{D}{c^2} - \frac{m^2}{r^2} -
\frac{2m\Omega}{r\tomega}
\left(\frac{d}{dr}\ln\frac{\Omega\Sigma}{D}\right) -
S^{1/2}\frac{d^2}{dr^2}S^{-1/2}\right]\eta =0~.   \label{perturbeq2}
\ee
This is our basic working equation.

\section{Propagation Diagram and Local Solutions Near Resonances}

Consider local free wave solution of the form
\be
\delta h\propto \exp\left[i\int^r\!k_r(s) ds\right]~.
\ee
For $|k_rr|\gg m$ and away from the $D=0$ region, 
we find from equation (\ref{perturbeq2})
\be
k_r^2+{D\over c^2}+
\frac{2m\Omega}{r\tomega}\left(\frac{d}{dr}\ln\frac{\Omega\Sigma}{D}\right)
\simeq 0.
\label{eq:disper}\ee
This is the general WKB dispersion relation.
Away from the region where $\tomega=0$, this reduces to the well-known
result $k_r^2\simeq -D/c^2$
[equation (\ref{eq:disp})]; in the vicinity of $\tomega=0$
this describes local Rossby waves, with 
[see equations (\ref{eq:disp2})-(\ref{eq:zeta})]
\be
\tomega\simeq {2m\Omega\over r k_r^2}
\left({d\ln\zeta\over dr}\right)_c.
\ee

Before studying global solutions to the wave equation (\ref{perturbeq1})
or (\ref{perturbeq2}), it is useful to consider the special resonant
locations in the disc. These can be
recognized by investigating the singular points and turning
points of the wave equation (\ref{perturbeq1}) or
(\ref{perturbeq2}), or by examining the characteristics of 
the dispersion relation (\ref{eq:disper}). 
The special radii are

(i) Lindblad resonances (LRs), where $D=0$ or $\tomega ^2=\kappa^2$,
including the outer Lindblad resonance (OLR) at $\tomega=\kappa$ and the inner
Lindblad resonance (ILR) at $\tomega=-\kappa$. The LRs are
apparent singularities of equation (\ref{perturbeq1})
or (\ref{perturbeq2}) -- all physical quantities
are finite at $D=0$. The LRs are turning points 
at which wave trains are reflected or transmitted. In the presence of an external force, waves are launched from LRs.

(ii) Corotation resonance (CR), where $\tomega=0$. In general,
the CR is a singular point of the wave equation
except in the special case of $d\zeta/dr=0$ at corotation. 
Some physical quantities (e.g., azimuthal velocity
perturbation) are divergent at corotation. Physically, this singularity
signifies that a steady emission or absorption of wave action
may occur there. 

From equation (\ref{perturbeq2}), we define the effective potential
for wave propagation by
\ba
V_{\rm eff}(r) &=&\frac{D}{c^2}+\frac{m^2}{r^2} +
\frac{2m\Omega}{r\tomega}
\left(\frac{d}{dr}\ln\frac{\Omega\Sigma}{D}\right)
+S^{1/2}\frac{d^2}{dr^2}S^{-1/2}\nonumber\\
&=&V_{\rm eff,0}(r)+\Delta V_{\rm eff}(r),
\ea
where
\ba
V_{\rm eff,0}(r) &=&\frac{D}{c^2}+\frac{m^2}{r^2} -
\frac{2m\Omega}{r\tomega} \left(\frac{d}{dr}\ln\zeta\right),\\
\Delta V_{\rm eff}(r) &=&
-\frac{2m\Omega}{r\tomega} \frac{d}{dr}\ln{D\over\kappa^2}
+S^{1/2}\frac{d^2}{dr^2}S^{-1/2}.
\ea
Clearly, wave propagation is possible only in the region where 
$V_{\rm eff}(r)<0$. Figures 1-3 depict the wave propagation diagrams 
for the cases of $(d\zeta/dr)_c=0,~<0$ and $>0$, respectively.
We are interested in the parameter regime $c/(r\Omega)\ll 1$ and $m$ is
of order unity. Note that the apparent singularity in $\Delta V_{\rm eff}(r)$
at $D=0$ can be eliminated by analysing the wave solution around the LRs 
(see section 3.1 below). Thus we also show $V_{\rm eff,0}(r)$ in Figs.~1-3.

We now consider the behaviors of the perturbations around the LRs and CR.

\begin{figure}
\centering
\includegraphics[width=13cm]{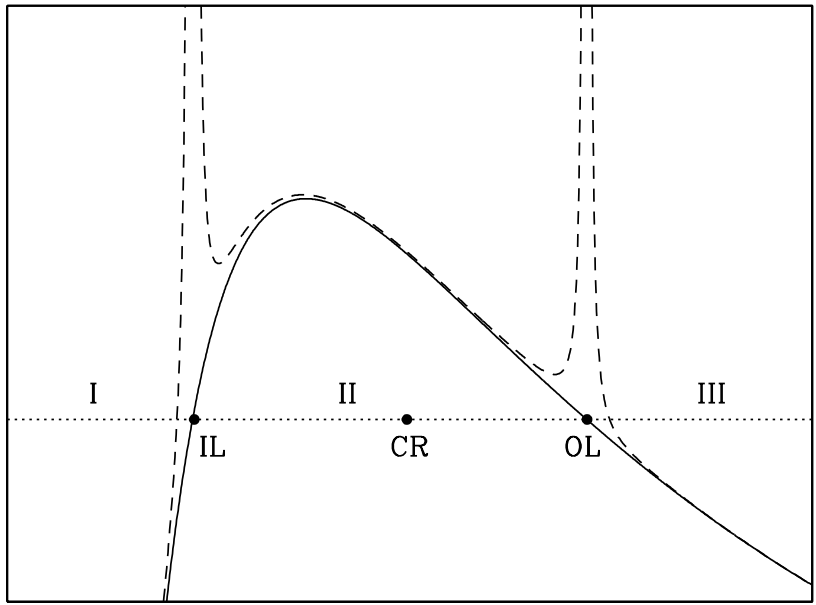}
{\vskip -3cm}
\caption{Wave propagation diagram in Keplerian discs:
A sketch of the effective potential $V_{\rm eff,0}(r)$ 
(solid line) and $V_{\rm eff}(r)$ (dashed line) as a function of $r$,
for the case of $(d\zeta/dr)_c=0$.
Waves can propagate only in the region where $V_{\rm eff}(r)< 0$, i.e., 
where the curves are below the dotted line. 
The three special locations are denoted by 
IL (Inner Lindblad Resonance), OL (Outer Lindblad Resonance) and CR 
(Corotation Resonance).
The divergence in the $V_{\rm eff}(r)$ curve around IL and OL represents
an apparent singularity.} 
\label{fig1}
\end{figure}	

\begin{figure}
\centering
\includegraphics[width=13cm]{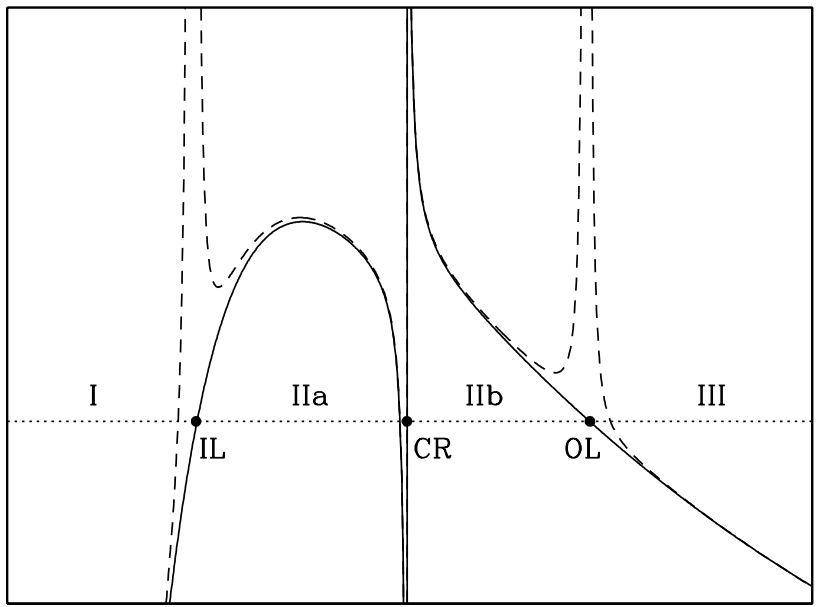}
{\vskip -3cm}
\caption{Same as Fig.~1, except for the case of 
negative vortensity gradient, $(d\zeta/dr)_c<0$ (or 
$\nu<0$). Note the CR represents a singularity, and the Rossby wave
zone lies inside the corotation radius.}
\label{fig2}
\end{figure}

\begin{figure}
\centering
\includegraphics[width=13cm]{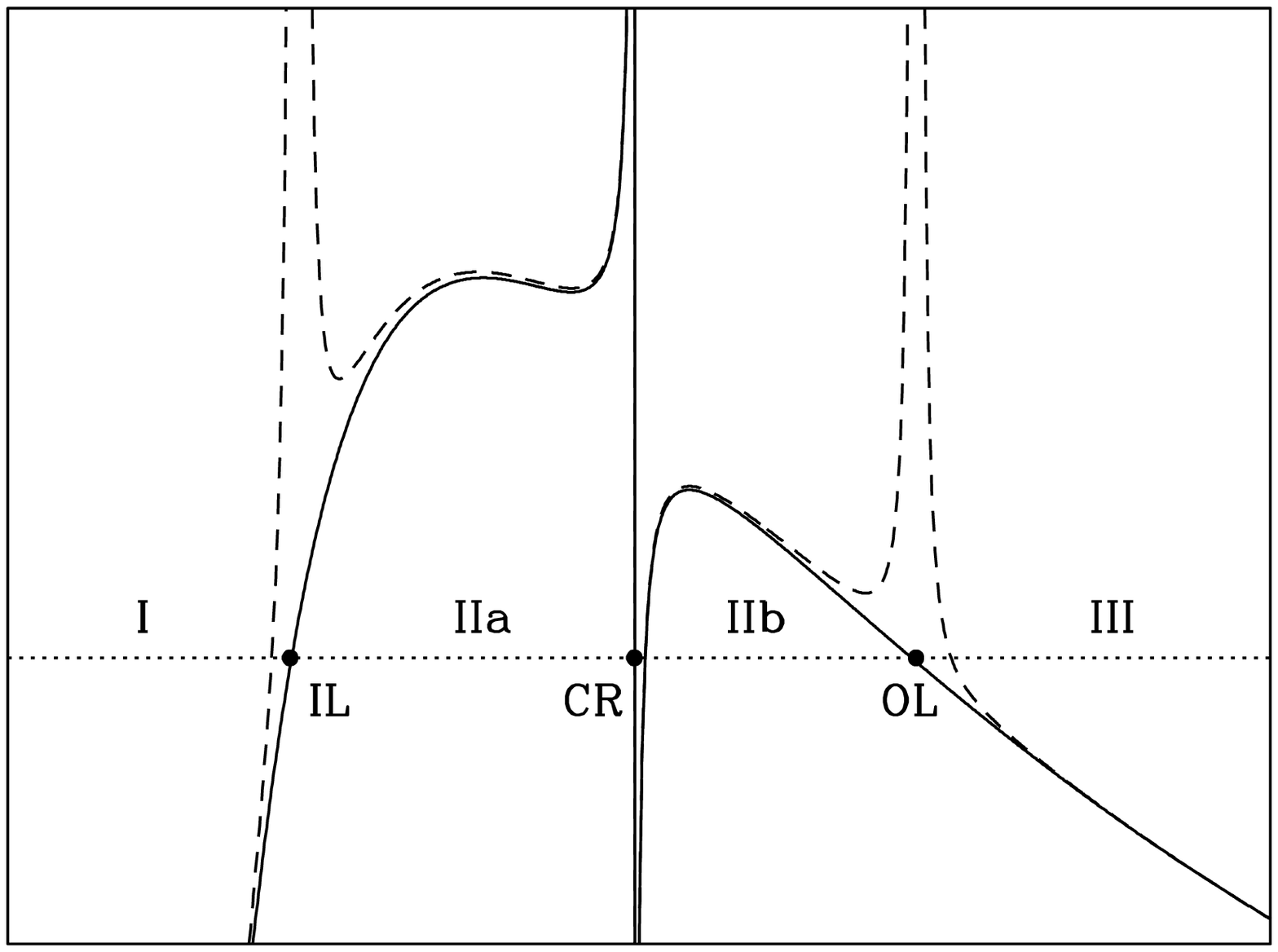}
{\vskip -3cm}
\caption{Same as Fig.~1, except for the case of 
positive vortensity gradient, $(d\zeta/dr)_c>0$ (or 
$\nu>0$), for which the Rossby wave
zone lies outside the corotation radius.}
\label{fig3}
\end{figure}

\subsection{Solution Around Lindblad Resonances}

Equation \eqref{perturbeq2} has an apparent singularity at the LRs,
where $D \rightarrow 0$. For concreteness we will 
explicitly examine the outer Lindblad resonance (OLR);
a similar solution can be
found for the inner Lindblad resonance (ILR).

In the vicinity of the OLR, equation \eqref{perturbeq2} becomes
\be
\frac{d^2}{dr^2} \eta + \left(k^2 - k \frac{d^2}{dr^2} \frac{1}{k}
+{4m\Omega\over r\tomega k}{dk\over dr}\right)\eta = 0, \label{LRdiffeq}
\ee
where $k^2\equiv -D/c^2$.
The last term inside $(\cdots)$ is smaller than the second term and will be
neglected. Changing the independent variable from $r$ to the dimensionless 
integrated phase 
\be
z = \int_{r_{\rm OL}}^r k\,dr,
\ee
we have 
\be
V'' + \left[ \frac{k''}{2k} - \frac{3}{4}\left(\frac{k'}{k}\right)^2
+ 1\right]V = 0,
\label{Vequation}\ee
where 
\be 
V= \sqrt{k}\,\eta=\sqrt{k\over S}\delta h,
\ee
and the prime denotes differentiation with respect to $z$.
Note that near OLR, $k^2 \simeq C(r-r_{\rm OL})$, with  
$C=(-c^{-2}dD/dr)_{\rm OL}>0$ a constant, we have
\be
z =
\Biggl\{
\begin{array}{ll}
\frac{2}{3}C^{1/2}(r-r_{OL})^{3/2} &\qquad \qquad \textrm{for }
r > r_{\rm OL} \\
 \frac{2}{3} C^{1/2}(r_{\rm OL}-r)^{3/2}e^{i3\pi/2}
&\qquad \qquad \textrm{for } r < r_{\rm OL}\end{array}
\ee
We can then express $k$ in terms of $z$ as
\be 
k = \left(\frac{3}{2} C\right)^{1/3} z^{1/3}~.\label{kintermsofy}
\ee
Using equation \eqref{kintermsofy} in equation \eqref{Vequation} we have
\be
V'' + \left( 1 - \frac{7}{36z^2}\right) V = 0 \label{finalVeqn}~.
\ee

Equation (\ref{finalVeqn}) has two independent solutions in terms of Bessel
function (Abramowitz \& Stegun 1964)
\be
V = \sqrt{z} J_{\pm 2/3} (z) 
= \frac{1}{(12 z)^{1/6}}\left[\pm\sqrt{3}{\rm Ai}'(-Z)+{\rm Bi}'(-Z)\right]~,  
\ee
where $Z=(3z/2)^{2/3}$, and ${\rm Ai}',~{\rm Bi}'$ are the derivatives of 
the Airy functions. Instead of using $\sqrt{z}J_{\pm 2/3}(z)$, we 
we can construct two linearly independent solutions for $\eta$ in a form 
convenient for asymptotic matching:
\be
\eta_1 = -\left({\pi\over k}\right)^{1/2}\!\left({2\over 3z}\right)^{1/6}\!
{\rm Ai}'(-Z)\sim 
\Biggl\{\begin{array}{ll}
\frac{1}{\sqrt{k}}\cos\left ( z + \pi/4\right) & \quad \textrm{for } 
|z| \gg 1\textrm{ and }\arg(z) = 0\\
\frac{1}{2\sqrt{k}}\exp\left(-|z|\right) & \quad \textrm{for }|z| \gg 1
\textrm{ and }\arg(z) = 3\pi/2
\end{array} 
\ee
\be
\eta_2 = \left({\pi\over k}\right)^{1/2}\!\left({2\over 3z}\right)^{1/6}\!
{\rm Bi}'(-Z)\sim 
\Biggl\{\begin{array}{ll}
\frac{1}{\sqrt{k}}\sin\left( z + \pi/4\right) & \quad \textrm{for } 
|z| \gg 1\textrm{ and }\arg(z) = 0\\
\frac{1}{\sqrt{k}}\exp\left(|z|\right) & \quad \textrm{for }|z| 
\gg 1\textrm{ and }
\arg(z) = 3\pi/2\end{array}  
\ee
where $\sim$ indicates asymptotic expansions.
This gives the connection formulae for the enthalpy perturbation
at the OLR\footnote{The usage here of $\gg$ is used as a shorthand for the range of validity for an asymptotic expansion of a local solution. The fitting formulae are to be used far enough away from the resonances so that the asymptotic expansion is valid, but close enough that the local approximation made in (22) holds.}:
\be
\delta h_1 \sim
\Biggl\{\begin{array}{ll}
\frac{1}{2}\sqrt{S/k}\, \exp\left(-\int_r^{r_{\rm OL}}\!|k| \,dr \right) 
& \qquad \textrm{for } r \ll r_{\rm OL}\\
\sqrt{S/k}\, \cos\left(\int_{r_{\rm OL}}^r\! k\, dr + \pi/4\right) 
&\qquad \textrm{for } r\gg r_{\rm OL}
\end{array}
\label{eq:con1}\ee
\be
\delta h_2  \sim
\Biggl\{\begin{array}{ll}
\sqrt{S/k} \exp\left(\int_r^{r_{\rm OL}}\! |k|\, dr \right) 
& \qquad \textrm{for } r \ll r_{OL}\\
\sqrt{S/k}\, \sin\left(\int_{r_{\rm OL}}^r\! k\, dr + \pi/4\right) 
&\qquad \textrm{for } r\gg r_{\rm OL}
\end{array}
\label{eq:con2}\ee

The connection formulae for ILR can be similarly derived:
\footnote{Note that for the ILR, $k$ is real for $r <  r_{\rm IL}$ 
and imaginary for $r > r_{\rm IL}$, while for 
the OLR, $k$ is real for $r >  r_{\rm OL}$ and imaginary for 
$r < r_{\rm OL}$.}
\be
\delta h_1 \sim
\Biggl\{\begin{array}{ll}
\frac{1}{2}\sqrt{S/k}\, \exp\left(-\int^r_{r_{\rm IL}}\!|k| \,dr \right) 
& \qquad \textrm{for } r \gg r_{\rm IL}\\
\sqrt{S/k}\, \cos\left(\int^{r_{\rm IL}}_r\! k\, dr + \pi/4\right) 
&\qquad \textrm{for } r\ll r_{\rm IL}
\end{array}
\label{eq:conn1}\ee
\be
\delta h_2  \sim
\Biggl\{\begin{array}{ll}
\sqrt{S/k} \exp\left(\int^r_{r_{\rm IL}}\! |k|\, dr \right) 
& \qquad \textrm{for } r \gg r_{IL}\\
\sqrt{S/k}\, \sin\left(\int^{r_{\rm IL}}_r\! k\, dr + \pi/4\right) 
&\qquad \textrm{for } r\ll r_{\rm IL}
\end{array}
\label{eq:conn2}\ee

\subsection{Solution Around Corotation Radius}

In the vicinity of the corotation radius $r_c$, we can rewrite \eqref{perturbeq2} dropping the $m^2/r^2$ and $S$ terms compared to the singular term proportional to $1/\tomega$, giving
\be
\left[\frac{d^2}{dr^2} - \tilde{k}^2 
+{2\over q}\left({d\over dr}\ln{\kappa^2\over\Omega\Sigma}\right)_c
{1\over r-R_c}\right]\eta = 0,
\label{pnearcr}\ee
where 
\be
\tilde{k}^2 \equiv \frac{D}{c^2},\quad
q\equiv -\left({d\ln \Omega\over d\ln r}\right)_c,\quad
R_c\equiv r_c-i{r_c\omega_i\over q\omega_r}.
\ee
Here we have introduced a small imaginary part to the wave frequency, so that
$\omega=\omega_r+i\omega_i$. To study the response of the disc to a slowly 
increasing perturbation, we require $\omega_i>0$. 
Defining 
\be
x = \int_{r_{c}}^r 2\tilde{k}\,dr, \quad \eta = \frac{1}{\sqrt{\tilde{k}}}\psi,
\ee
and recognizing that $\tilde k$ can be treated as a constant around corotation,
we have
\be 
\frac{d^2}{dx^2} \psi + \left(
-\frac{1}{4} + \frac{\nu}{x+i\epsilon}\right)\psi =0
\label{Whittakerdiffeq} \ee 
where $\epsilon=2{\tilde k}r_c\omega_i/(q\omega_r)$ and
\be
\nu={1\over q\tilde k}\left({d\over dr}\ln{\kappa^2\over\Omega\Sigma}\right)_c
=\left({c\over q\kappa}{d\over dr}\ln\zeta\right)_c. \label{eq:nu}
\ee
In equation 
(\ref{Whittakerdiffeq}) $\epsilon > 0$, consistent with the initial value 
problem in which the perturbation is gradually turned on starting 
from $t=-\infty$. The parameter $\nu$ determines the width of the Rossby wave region, 
$\Delta r_R = (2c/\kappa)|\nu|$. For $d\ln\zeta/dr\sim 1/r$, we have
$|\nu|\sim (H/r)_c$ ($H$ is the disc scale height). \footnote{Note that the WKB wavenumber $k_{r}$ in the Rossby region ranges from $\infty$ (at $r=r_c$) to $H^{-1}$ (at $r\sim r_c \pm \Delta r_R$) [see equation (18)] while the width of the region is of order $H^2/r$ for most Keplerian discs. Free Rossby waves tend to be sheared away by differential rotation (Tagger 2001).}

Equation (\ref{Whittakerdiffeq}) is the differential equation for the 
Whittaker function in complex variable $z = x +
i\epsilon$ with index $1/2$ (Abramowitz \& Stegun 1964).
The two linearly independent solutions convenient for the connection are
\be 
\psi_- = \textrm{W}_{\nu,1/2} (z), \qquad 
\psi_+ = e^{-i\pi\nu}\textrm{W}_{-\nu, 1/2} (ze^{-i\pi}) + 
\frac{1}{2} T_0 \textrm{W}_{\nu,1/2} (z), 
\label{eq:psi}\ee
where $T_0$ is the Stokes multiplier defined 
below, and $z$ 
is defined in the complex plane so that $\arg(z)$ ranges from $0$ to $\pi$.
The particular linear combinations of Whittaker functions in (\ref{eq:psi})
are chosen so that appropriate asymptotic expansions can be obtained.
To obtain these asymptotic expansions and the 
connection formulae around the corotation, one must carefully consider 
the Stokes phenomenon, which alters the form of the asymptotic expansion 
of a function depending on the position of $z$ in the complex
plane. Since the appropriate asymptotic expansions (which connect 
the solution $W_{\pm\nu,1/2}$ analytically in different regions of the 
complex plane) are not readily available, we relegate the discussion of the 
Stokes phenomenon for the Whittaker function around the corotation 
to Appendix A\footnote{The Stokes phenomenon is present 
in our Lindblad resonance analysis as well, as we see, for example, 
the asymptotic expansion for $Bi'(-(3z/2)^{2/3})$ takes the form 
$e^{+|z|}$ for $\arg(z) = 3\pi/2$ but $\sin(z + \pi/4)$ 
(as opposed to $e^{iz}$) for $\arg(z) =0$. 
The resulting connection formulae are well known for the Airy functions.}.
The resulting connection formulae are
\ba
\delta h_- &\sim&
\Biggl\{\begin{array}{ll}
\sqrt{S/k}\, \exp\left(-\int_{r_c}^{r}\! \tilde k \,dr \right) 
& \qquad \qquad \qquad ~~\textrm{for } r \gg r_{c}\\
\sqrt{S/k}\, e^{i\pi\nu}\exp\left(+\int_r^{r_{c}}\! \tilde k \,dr \right) 
+ \sqrt{S/k}\, \frac{T_1}{2} e^{-i\pi\nu} \exp\left(-\int_r^{r_{c}}\! \tilde k \,dr \right) 
&\qquad \qquad \qquad~~\textrm{for } r\ll r_{c}~.
\end{array}\label{whitaker1}\\
\delta h_+ &\sim&
\Biggl\{\begin{array}{ll}
\sqrt{S/k}\, \exp\left(+\int_{r_c}^{r}\! \tilde k \,dr \right) 
& \qquad \textrm{for } r \gg r_{c}\\
\sqrt{S/k}\, \frac{T_0}{2} e^{i\pi\nu}\exp\left(+\int_r^{r_{c}}\! \tilde k \,dr \right) 
+ \sqrt{S/k}\, \left(1 + \frac{T_1T_0}{4}\right)  e^{-i\pi\nu}\exp\left(-\int_r^{r_{c}}\! \tilde k \,dr \right) 
&\qquad \textrm{for } r\ll r_{c}
\end{array}\label{whitaker2}
\ea
where $T_0$ and $T_1$ are the Stokes multipliers, 
\be 
T_0={2\pi i\over\Gamma(\nu)\Gamma(1+\nu)},\qquad
T_1={2\pi i\,e^{i2\pi\nu}\over\Gamma(-\nu)\Gamma(1-\nu)}.
\ee
Note that for $|\nu| \ll 1$, 
$[\Gamma(\pm \nu)]^{-1}=\pm\nu +\gamma\nu^2+\cdots$, 
where $\gamma=0.5772$ is the Euler constant.

\section{Global WKB Solutions and Calculation of Reflectivity}

In this section we consider a wave train which approaches corotation 
from small radii ($r\ll r_{\rm IL}$). Its propagation is impeded 
by the potential barrier between $r_{\rm IL}$ and $r_{\rm OL}$.
The incident wave is partially transmitted beyond the OLR and a reflected
wave propagates from the ILR toward small radii.
We will derive the explicit expressions for the (complex) reflection
coefficient ${\cal R}$ and transmission coefficient ${\cal T}$.

From the dispersion relation [equation (\ref{eq:disp}), or equation 
(\ref{eq:disper})
away from corotation], we find that the radial group velocity of the waves is 
\be
c_g={d\omega\over dk_r}={k_rc^2\over\tomega (1-\kappa^2/\tomega^2)}.
\ee
Thus the sign of $c_g/c_p$ (where $c_p=\omega/k_r$ is the phase velocity)
is positive for $r>r_{\rm OL}$ and negative for $r<r_{\rm IL}$.
This implies that in the $r>r_{\rm OL}$ region, the outgoing (transmitted) 
wave has the form $\exp(i\int^r k_r\,dr)$ (assuming $k_r>0$). 
In the $r<r_{\rm IL}$ region, the incident wave (propagating from small 
radii toward corotation) has the form $\exp(-i\int^r k_r\,dr)$, while
the reflected wave has the form $\exp(i\int^r k_r\,dr)$.

A well-known property of density waves is that for $r<r_{\rm IL}$
the wave carries negative energy (or angular momentum), 
while for $r>r_{\rm OL}$
the wave carries positive energy. An incident
wave $\exp(-i\int^r k_r\,dr)$, carrying energy of the amount $(-1)$, 
will give rise to a reflected wave ${\cal R}\exp(i\int^r k_r\,dr)$ 
and a transmitted wave ${\cal T}\exp(i\int^r k_r\,dr)$. Let ${\cal D}_c$ 
be the energy dissipated at the corotation. Then energy conservation gives
$-1=(-1)|{\cal R}|^2+|{\cal T}|^2+{\cal D}_c$, or
\be
|{\cal R}|^2=1+|{\cal T}|^2+{\cal D}_c. \label{eq:Econs}
\ee
Because of the singularity at corotation and the associated 
energy absorption, we first consider the simple case where
the corotation singularity is neglected (section 4.1)
before examining the general case (section 4.2).

\subsection{Neglecting Corotation Singularity: Corotation Amplifier}

Here we consider the case where the vortensity has zero slope
at corotation, i.e., $(d\zeta/dr)_c=0$. This would occur for disc models
where the specific angular momentum $r^2\Omega$ is constant (as in the 
original Papaloizou \& Pringle 1984 analysis), or for shearing sheet
approximation (as in Narayan et al.~1987). In this case, there is no
corotation singularity and no absorption of wave energy, and we always 
obtain super-reflection
\be
|{\cal R}|^2=1+|{\cal T}|^2>1.
\ee
This is the essence of the corotation amplifier.

To derive the complex ${\cal R}$ and ${\cal T}$, we assume
that the outgoing wave in the $r> r_{\rm OL}$ region
(region III in Fig.\ref{fig1}) is given by [see 
eqs.~(\ref{eq:con1})-(\ref{eq:con2})]
\be
\delta h =\sqrt{S/k}\exp\left( i \int_{r_{\rm OL}}^r k\,dr
+ \frac{\pi}{4}\right),
\label{eq:outside}\ee
where $k^2\equiv -D/c^2$.
The connection formulae (\ref{eq:con1})-(\ref{eq:con2})
then give for the evanescent zone (region II in Fig.~1):
\ba
\delta h &\simeq&  \frac{\sqrt{S/k}}{2}\exp\left(-\int_r^{r_{\rm OL}}|k|\,dr 
\right) + i\sqrt{S/k}\exp\left(\int_r^{r_{\rm OL}} |k| dr\right) \nonumber\\
&=& \frac{\sqrt{S/k}}{2}
\exp(-\Theta_{\rm{II}}) 
\exp\left(\int_{r_{\rm IL}}^{r}|k| dr \right) + i\sqrt{S/k}
\exp(+\Theta_{\rm{II}})\exp\left(-\int_{r_{\rm IL}}^r |k| dr\right) 
\ea
where 
\be
\Theta_{\rm{II}} = \int_{r_{\rm IL}}^{r_{\rm OL}} |k|\, dr
= \int_{r_{\rm IL}}^{r_{\rm OL}} {\sqrt{|D|}\over c}\, dr.
\ee
Using the connection formulae at ILR [eqs.~(\ref{eq:conn1})-(\ref{eq:conn2})],
we find that for $r < r_{\rm IL}$ (region I in Fig.~1)
\be
\delta h \simeq \frac{\sqrt{S/k}}{2}e^{-\Theta_{\rm{II}}}
\sin\left(\int_r^{r_{\rm IL}} k\,dr +\frac{\pi}{4}\right) + i2\sqrt{S/k}
e^{\Theta_{\rm{II}}}\cos\left(\int_r^{r_{\rm IL}} k\,dr 
+\frac{\pi}{4}\right)~.
\ee
Expressing this in terms of traveling waves, and defining 
$y = \int_{r_{\rm IL}}^r k dr - \pi/4$ we have
\be
\delta h \simeq i\sqrt{S/k} \left[ e^{-iy} \left(e^{+\Theta_{\rm{II}}} 
- \frac{1}{4}e^{-\Theta_{\rm{II}}}\right) + e^{+iy} \left(
e^{\Theta_{\rm{II}}} + \frac{1}{4}e^{-\Theta_{\rm{II}}}\right)\right].
\label{eq:inside}\ee
Thus the reflection coefficient is 
\be
{\cal R} = \frac{e^{\Theta_{\rm{II}}} + \frac{1}{4}e^{-\Theta_{\rm{II}}}}{e^{\Theta_{\rm{II}}} -
\frac{1}{4}e^{-\Theta_{\rm{II}}}}.\label{SimpleReflection}
\ee
Comparing equation (\ref{eq:inside}) with equation (\ref{eq:outside}), 
we obtain the transmission coefficient
\be
{\cal T}={-i\over e^{+\Theta_{\rm II}}-{1\over 4}e^{-\Theta_{\rm II}}}.
\label{Simpletrans}\ee
As expected, $|{\cal R}|^2=1+|{\cal T}|^2>1$.

\subsection{Including Corotation Singularity}

As noted before, for discs with nonzero vortensity gradient
($d\zeta/dr\ne 0$), the singularity at corotation implies the absorption of
wave energy (or angular momentum). Similar situations
occur in geophysical wave systems (Dickenson 1968) and for waves in plasmas
(Landau damping). 
In Appendix B, we discuss the toy problem of resonant tunneling which 
shares the similar energy absorption feature as the corotation singularity.
Previous works on global modes in disc tori
(e.g. Papaloizou \& Pringle 1987; Goldreich et al.~1986; Narayan et al.~1987)
suggest that the sign of $d\zeta/dr$ determines whether the singularity 
acts to stabilize or destabilize a global mode. 
Here we derive the explicit expression for 
the reflectivity and the related source term ${\cal D}_c$.

As in section 4.1, we assume an outgoing wave in Region III, 
and the connection formulae at the OLR then give for Region IIb of 
Fig.~\ref{fig2} or Fig.~\ref{fig3}
\be
\delta h \simeq
\frac{\sqrt{S/k}}{2}\exp(-\Theta_{\rm{IIb}})\exp\left(\int_{r_{c}}^r
|k|\,dr\right) + i\sqrt{S/k}\exp(+\Theta_{\rm{IIb}})\exp
\left(-\int_{r_{c}}^r |k|\,dr\right),
\label{eq:dhc}\ee
where 
\be
\Theta_{\rm{IIb}} = \int_{r_{c}}^{r_{\rm OL}} |k|\,dr.
\ee
Equation (\ref{eq:dhc}) is the asymptotic solution away from the $r_c$ in
region IIb. The corresponding general solution around $r_c$ is
\be
\delta h = \frac{\sqrt{S/k}}{2}\exp(-\Theta_{\rm IIb})\psi_+(r) + i\sqrt{S/k}\exp(\Theta_{\rm IIb}) \psi_-(r),
\label{eq:general}\ee
where $\psi_+$ and $\psi_-$ are given by (\ref{eq:psi}).
Using equations \eqref{whitaker1} and \eqref{whitaker2} to match asymptotes 
over the corotation singularity, we obtain in Region $\rm{IIa}$
\ba
\delta h &\simeq& \left[\frac{1}{2}\exp(-\Theta_{\rm{IIb}})(1+\frac{1}{4} T_0 T_1) 
+\frac{i}{2} T_1
\exp(+\Theta_{\rm{IIb}})
\right]\sqrt{S/k}~e^{-i\pi\nu}\exp\left(-\int_r^{r_{c}}\!|k|\,dr \right)\nonumber\\
&\qquad& +i\Bigl[
\exp(+\Theta_{\rm{IIb}})
-\frac{i}{4} T_0 \exp(-\Theta_{\rm IIb})\Bigr]
\sqrt{S/k} ~ e^{i\pi\nu}\exp\left(\int_r^{r_{c}}|k|\,dr\right).
\ea
Using the connection formulae at the ILR, we have for Region I:
\ba
\delta h &\simeq & \left[\frac{1}{2}\exp(-\Theta_{\rm{II}}) 
\left(1+\frac{1}{4} T_0 T_1\right)
 + \frac{i}{2} T_1\exp(\Theta_{\rm{IIb}}-\Theta_{\rm{IIa}}) \right]
\sqrt{S/k}~e^{-i\pi\nu}\sin\left(\!\int_r^{r_{\rm IL}}\!\!k\,dr 
+\frac{\pi}{4}\right)\nonumber\\ 
&\quad & + 2i \left[ \exp(+\Theta_{\rm{II}}) 
-\frac{i}{4} T_0 \exp(\Theta_{\rm{IIa}}
-\Theta_{\rm{IIb}}) \right]\sqrt{S/k}~e^{i\pi\nu}\cos\left(\!\int_r^{r_{\rm IL}}\!\!k\,dr
+\frac{\pi}{4}\right)\nonumber\\
&=& \left[1+{1\over 4}\,e^{-i2\pi\nu}e^{-2\Theta_{\rm II}}\left(1+\frac{1}{4} T_0 T_1\right)
+{i\over 4} T_1\,e^{-i2\pi\nu}e^{-2\Theta_{\rm IIa}}
-{i\over 4} T_0 e^{-2\Theta_{\rm IIb}}\right]i\sqrt{S/k}\, 
e^{\Theta_{\rm II}}e^{i\pi\nu}e^{iy}\nonumber\\
& \quad & +\left[1-{1\over 4}\,e^{-i2\pi\nu}e^{-2\Theta_{\rm II}}\left(1+{1\over 4}  T_0  T_1\right)
-{i\over 4} T_1 e^{-i2\pi\nu}e^{-2\Theta_{\rm IIa}}
-{i\over 4} T_0 e^{-2\Theta_{\rm IIb}}\right]i\sqrt{S/k}\, 
e^{\Theta_{\rm II}} e^{i\pi\nu} e^{-iy},
\ea
where $y = \int_{r_{\rm IL}}^r k dr - \pi/4$ and
\be
\Theta_{\rm{IIa}} =  \int_{r_{\rm IL}}^{r_{c}}\!\!|k|\,dr,\qquad
\Theta_{\rm II}=\Theta_{\rm{IIa}}+\Theta_{\rm{IIb}}
=\int_{r_{\rm IL}}^{r_{\rm OL}}\!\!|k|\,dr.
\ee
The reflection coefficient and transmission coefficient are then
\ba
&&{\cal R}={ 1+{1\over 4}\,e^{-i2\pi\nu}e^{-2\Theta_{\rm II}}\left(1+{1 \over 4} T_0 T_1\right)
+{i\over 4} T_1 e^{-i2\pi\nu}e^{-2\Theta_{\rm IIa}}
-{i\over 4} T_0e^{-2\Theta_{\rm IIb}}
\over
1-{1\over 4}\,e^{-i2\pi\nu}e^{-2\Theta_{\rm II}}\left(1+{1 \over 4}  T_0 T_1\right)
-{i\over 4} T_1 e^{-i2\pi\nu}e^{-2\Theta_{\rm IIa}}
-{i\over 4} T_0 e^{-2\Theta_{\rm IIb}}},\nonumber\\
&& \quad = { 1 + {1 \over 4} \left(e^{-i 2\pi \nu} + \sin^2 \pi \nu \right)e^{-2\Theta_{\rm II}}
+ {\pi \nu \over 2} {e^{-2\Theta_{\rm IIa}}\over (\Gamma(1-\nu))^2}
- {\pi \nu \over 2} {e^{-2\Theta_{\rm IIb}}\over (\Gamma(1 + \nu))^2}
\over
1 - {1 \over 4} \left(e^{-i 2\pi \nu} + \sin^2 \pi \nu \right)e^{-2\Theta_{\rm II}}
- {\pi \nu \over 2} {e^{-2\Theta_{\rm IIa}}\over (\Gamma(1-\nu))^2}
-{\pi \nu \over 2} {e^{-2\Theta_{\rm IIb}}\over (\Gamma(1 + \nu))^2}
}
\label{eq:RwithGamma}\\
&&{\cal T}={-i\, e^{-\Theta_{\rm II}} e^{i\pi\nu}
\over
1-{1\over 4}\,e^{-i2\pi\nu}e^{-2\Theta_{\rm II}}\left(1+{1 \over 4} T_0 T_1\right)
-{i\over 4} T_1 e^{-i2\pi\nu} e^{-2\Theta_{\rm IIa}}
-{i\over 4}T_0 e^{-2\Theta_{\rm IIb}}}\nonumber\\
&&\quad =
{- i e^{-\Theta_{\rm II}}e^{i\pi\nu}
\over
1 - {1 \over 4} \left(e^{-i 2\pi \nu} + \sin^2 \pi \nu \right)e^{-2\Theta_{\rm II}}
- {\pi \nu \over 2} {e^{-2\Theta_{\rm IIa}}\over (\Gamma(1-\nu))^2}
-{\pi \nu \over 2} {e^{-2\Theta_{\rm IIb}}\over (\Gamma(1 + \nu))^2}
}
\label{eq:trans}\ea
The dissipation term due to the corotation singularity obtained from
${\cal D}_c=|{\cal R}|^2 - 1- |{\cal T}|^2$ is
\be
{\cal D}_c = {
{\pi \nu \over 2} {\cos^2\pi\nu\over (\Gamma(1 + \nu))^2} e^{-2\Theta_{\rm II} - 2\Theta_{\rm IIb}} 
+ {2\pi \nu \over (\Gamma(1-\nu))^2} e^{-2\Theta_{\rm IIa}}
 \over |1 - {1 \over 4} \left(e^{-i 2\pi \nu} + \sin^2 \pi \nu \right)e^{-2\Theta_{\rm II}}
- {\pi \nu \over 2} {e^{-2\Theta_{\rm IIa}}\over (\Gamma(1-\nu))^2}
-{\pi \nu \over 2} {e^{-2\Theta_{\rm IIb}}\over (\Gamma(1 + \nu))^2}
|^2} .
\ee

For $|\nu| \ll 1$ 
equations (\ref{eq:RwithGamma}) and (\ref{eq:trans}) can be simplified,
and we have
\ba
&&{\cal R}\rightarrow { e^{\Theta_{\rm II}} + {1 \over 4} e^{-\Theta_{\rm II}}
\over { e^{\Theta_{\rm II}} - {1 \over 4} e^{-\Theta_{\rm II}}}} + { e^{+2\Theta_{\rm IIb}}-{1\over 4}  e^{-2\Theta_{\rm IIb}} \over \left(e^{\Theta_{\rm II}} - {1 \over 4} e^{-\Theta_{\rm II}}\right)^2} \pi \nu + {\cal O}[\nu^2] \label{eq:R},\label{eq:eqr}\\
&&{\cal T}\rightarrow { -i
\over { e^{\Theta_{\rm II}} - {1 \over 4} e^{-\Theta_{\rm II}}}} 
- {i \over 2} { e^{\Theta_{\rm IIb}-\Theta_{\rm IIa}}-e^{\Theta_{\rm IIa}-\Theta_{\rm IIb}} \over \left(e^{\Theta_{\rm II}} - {1 \over 4} e^{-\Theta_{\rm II}}\right)^2} \pi \nu + {\cal O}[\nu^2],\label{eq:eqt}\\
&&{\cal D}_c\rightarrow  {2 \left( e^{\Theta_{\rm II}} + {1\over 4} e^{-\Theta_{\rm II}}\right)
\left( e^{2\Theta_{\rm IIb}}-{1 \over 4} e^{-2\Theta_{\rm IIb}}\right) 
+ \left( e^{\Theta_{\rm IIb}-\Theta_{\rm IIa}}-e^{\Theta_{\rm IIa}-\Theta_{\rm IIb}}\right)
\over \left(e^{\Theta_{\rm II}} - {1 \over 4} e^{-\Theta_{\rm II}}\right)^3}
\pi \nu + {\cal O}[\nu^2]. \label{eq:S} 
\ea
Clearly, for $\nu=0$, equations (\ref{eq:eqr})-(\ref{eq:eqt}) reduce to
(\ref{SimpleReflection}) and (\ref{Simpletrans}).
 
\section{Wave Damping at Corotation: Alternative Calculation}

In the previous section we obtained the expression for the
dissipation term ${\cal D}_c$ at the corotation resonance using 
the reflection and transmission coefficients. Here we provide
a more direct derivation of ${\cal D}_c$ using the change of angular momentum
flux across the corotation radius.

In the absence of self-gravity, the angular momentum 
flux carried by the the waves in the disc is entirely due to advection.
The time-averaged transfer rate of the $z$-component of angular momentum 
across a cylinder of radius $r$ (in the outward direction) is given by 
(e.g. Goldreich \& Tremaine 1979)
\be
F(r)=r^2\Sigma(r)\int_0^{2\pi}\!d\phi\,{\rm Re}[\delta u_r(r,\phi,t)]
{\rm Re}[\delta u_\phi(r,\phi,t)].
\ee
Using equations (\ref{eq:ur})-(\ref{eq:uphi}) to express $\delta u_r$
and $\delta u_\phi$ in terms of $\delta h$, this reduces to 
(see Tanaka et al.~2002; Zhang \& Lai 2006)
\be
F(r) = \frac{\pi m r \Sigma}{D}{\rm Im}\left(\delta h\frac{d\delta
h^*}{dr} \right) .  
\ee

In Region III (see Figs.~1-3) the outgoing wave has the enthalpy 
perturbation given by (up to a proportional constant)
\be
\delta h = \sqrt{S/k}\, {\cal T} \exp \left( i \int^r_{r_{\rm OL}} k dr
+{\pi\over 4} \right)
\label{eq:ddh}\ee
Calculating the angular momentum flux (setting the incoming wave flux to 1),
\be
F(r\gg r_{\rm OL})\simeq \pi m |{\cal T}|^2,
\ee
we see that angular momentum is transferred outwards (positive flux) 
since waves in $r>r_{\rm OL}$ carries positive angular momentum.
For Region I we have [up to the same proportional constant as in
(\ref{eq:ddh})]
\be
\delta h = \sqrt{S/k}\left[\exp \left( - i \int^r_{r_{\rm IL}} k dr 
+{\pi\over 4}\right) + {\cal R} \exp \left( i \int^r_{r_{\rm IL}} k dr
-{\pi\over 4}\right) \right],
\ee
which gives the angular momentum flux:
\be
F(r\ll r_{\rm IL}) \simeq \pi m (|{\cal R}|^2 - 1).
\ee
We see that the incident wave carries negative angular momentum outward,
and the reflected wave transfers positive angular momentum. The net angular 
momentum transfer is positive (in the outward direction) for $|{\cal R}|>1$.

Now consider the angular flux near the corotation radius, at $r=r_c^-$ 
(just inside corotation ) and at $r=r_c^+$ (just outside corotation).
Integrating equation~\eqref{pnearcr} across the singularity,
we find the discontinuity in the enthalpy perturbation derivatives:
\be
\frac{d\,\delta h}{dr}\bigg|_{r_c^+} - \frac{d\,\delta
h}{dr}\bigg|_{r_c^-} = {2\pi i\over q}\left({d\over dr}\ln\zeta\right)\delta h
\bigg|_{r_c}=2\pi\nu i {\kappa\over c}\delta h\bigg|_{r_c}.
\ee
Here we have chosen to integrate from $r_c^-$ to $r_c^+$ by going through the 
upper complex plane, to be consistent with the physical requirement of 
a gradually growing perturbation, turned on at $t = -\infty$. Thus the change in the angular momentum 
flux across the corotation is 
\be
\Delta F_c=F(r_c^+)-F(r_c^-)
= -\frac{2\pi^2 m r \Sigma\nu\kappa }{cD} |\delta h|^2\bigg|_{r_c}.
\ee
The wavefunction around $r_c$ is given by equation (\ref{eq:general})
multiplied by ${\cal T}$. 
Noting that
\ba
\psi_-(r_c) &=& {\rm W}_{\nu, 1/2}(0) = \frac{1}{\Gamma(1-\nu)},\\
\psi_+(r_c) &=& e^{-i\pi\nu}\textrm{W}_{-\nu, 1/2} (0) + \frac{1}{2} T_0 \textrm{W}_{\nu,1/2} (0) = \frac{\cos \pi\nu}{\Gamma(1 + \nu)} ,
\ea
where we have used the convenient identity $\Gamma(\nu)\Gamma(1 - \nu) =
\pi/\sin\pi\nu$. We can evaluate the enthalpy perturbation at the
corotation, giving
\be
\delta h(r_c) = \sqrt{S/k}\,{\cal T}\left[  {1\over 2} e^{-\Theta_{\rm IIb}} {\cos \pi \nu \over \Gamma(1 + \nu)} + i e^{\Theta_{\rm IIb}}{1 \over \Gamma(1-\nu)}\right].
\ee
The change in angular momentum flux across the corotation is then
\be
\Delta F_c= -\pi m |{\cal T}|^2\nu \left[ 
{\pi \over 2} {\cos^2\pi\nu \over (\Gamma(1 + \nu))^2}e^{-2\Theta_{\rm IIb}}
+ {2\pi \over (\Gamma(1-\nu))^2} e^{2\Theta_{\rm IIb}}
\right].
\ee
With $F(r\ll r_{\rm IL})=F(r_c^-)$, $F(r\gg r_{\rm OL})=F(r_c^+)$,
and thus $F(r\ll r_{\rm IL})=F(r\gg r_{\rm OL})-\Delta F_c$,
we find 
\be
{\cal D}_c = |{\cal T}|^2 \nu \left[ 
{\pi \over 2} {\cos^2\pi\nu \over (\Gamma(1 + \nu))^2}e^{-2\Theta_{\rm IIb}}
+ {2\pi \over (\Gamma(1-\nu))^2} e^{2\Theta_{\rm IIb}}
\right].
\label{eq:scc}\ee
This expression exactly agrees with ${\cal D}_c$ given in section 4.2.

\section{Results and Discussion}

The key new results of this paper consist of the analytical 
expressions for the reflection coefficient ${\cal R}$, transmission 
coefficient ${\cal T}$ and the dissipation term ${\cal D}_c$
when a wave impinges upon the corotation barrier from small radii.
These expressions, (\ref{eq:RwithGamma})-(\ref{eq:S}) and (\ref{eq:scc}),
can be applied to discs with generic rotation and surface density profiles.

\begin{figure}
\centering
\includegraphics[width=13cm]{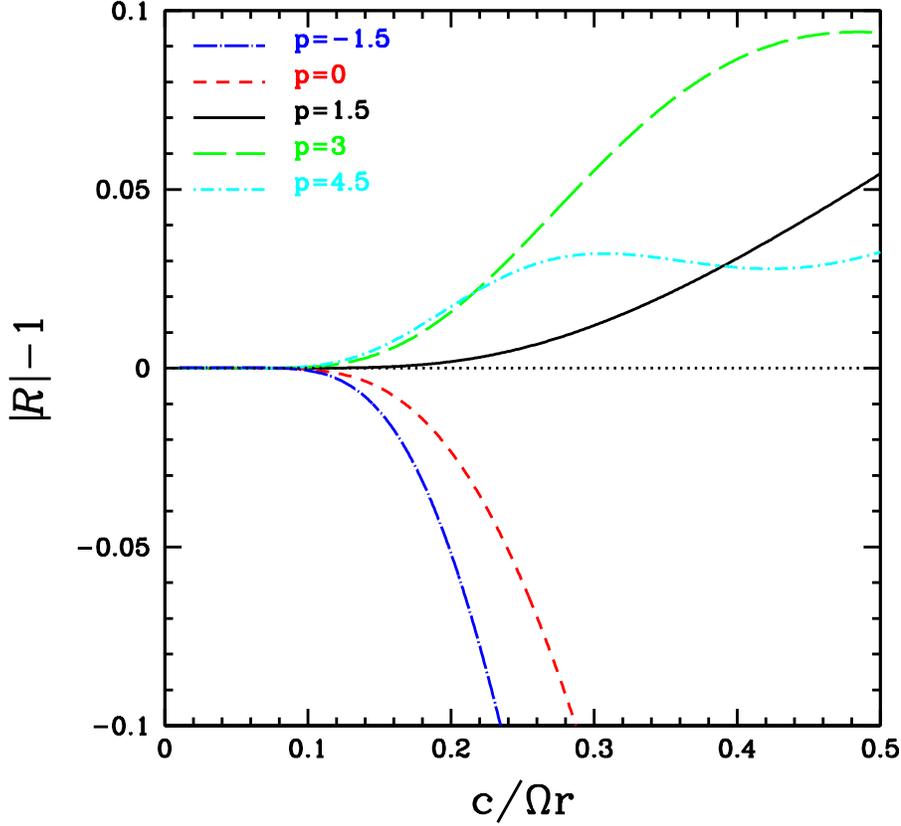}
{\vskip -1cm}
\caption{
The reflection coefficient as a function of $\beta=c/(r\Omega)$ for
Keplerian discs with surface density profile $\Sigma\propto r^{-p}$.
Note that for $p=1.5$, wave absorption at the corotation resonance is 
absent.
}
\label{fig4}
\end{figure}

\begin{figure}
\centering
\includegraphics[width=13cm]{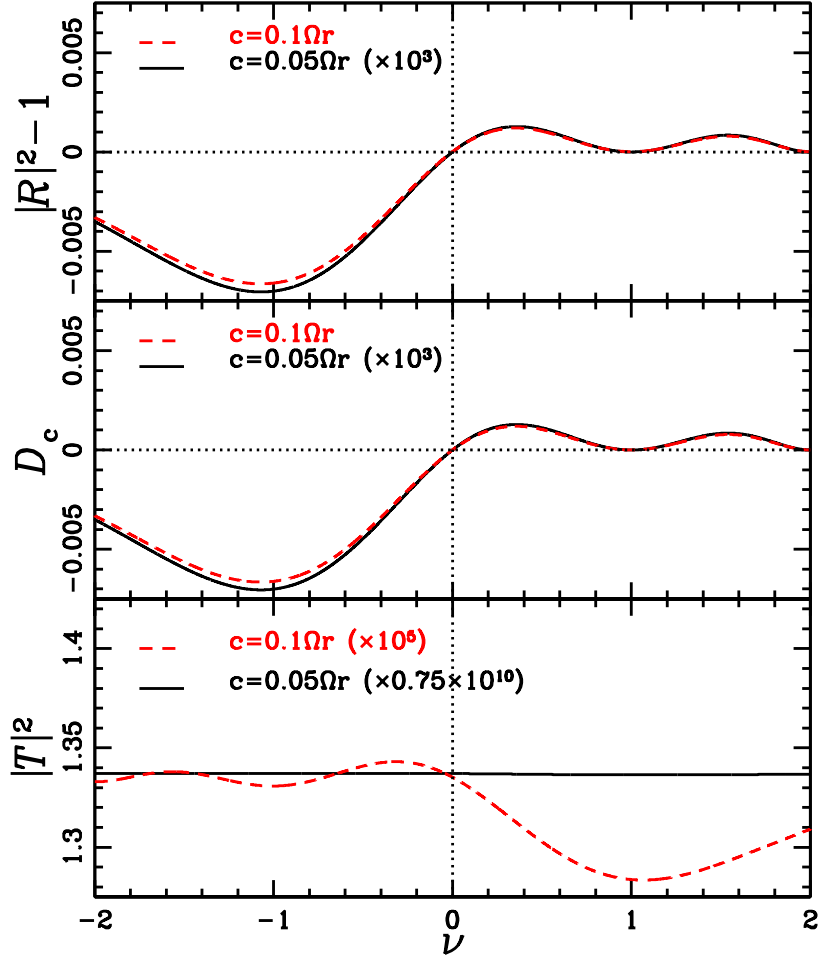}
\caption{The reflection coefficient, wave damping coefficient and
transmission coefficient as a function of $\nu$ for two different values
of $\beta=c/(r\Omega)$. Note that for $\nu=0$ (no corotation resonance),
wave damping is zero (${\cal D}_c=0$) 
and $|{\cal R}|^2-1$ is positive.
}
\label{fig5}
\end{figure}

For definiteness, here we illustrate our results using a (Newtonian) 
Keplerian disc model with 
\be
\Omega=\kappa\propto r^{-3/2},\quad
\Sigma\propto r^{-p},\quad 
{c\over r\Omega}=\beta,
\ee
where $p$ and $\beta$ are constants. The important parameter that
determines the behavior of the corotation singularity is [see eq.~
(\ref{eq:nu})]
\be
\nu=\left({2c\over 3\kappa}{d\over dr}\ln\zeta\right)_c=
{2\over 3}\beta\left(p-{3\over 2}\right).
\ee
Clearly the models are scale-free, and ${\cal R}$, ${\cal T}$ 
and ${\cal D}_c$ depend only on the two parameters $p$ and $\beta$
(or $\nu$). Figure 4 depicts $|{\cal R}|$ as a function of $\beta$ for 
different values of $p$, while Figure 5 shows $|{\cal R}|,~{\cal D}_c$ 
and $|{\cal T}|$ as a function of $\nu$ for $\beta=0.05$ and $\beta=0.1$.

A key result of paper is that wave absorption at the corotation resonance
plays an important role in determining the 
reflection and transmission of waves across the corotation barrier.
Without corotation resonance (as for discs with zero 
vortensity gradient, or $\nu=0$), super-reflection is
always achieved, but $|{\cal R}|^2-1\simeq \exp(-2\Theta_{\rm II})$
(assuming $\Theta_{\rm II}\gg 1$, where 
$\Theta_{\rm II}=\Theta_{\rm IIa}+\Theta_{\rm IIb}=
\int_{r_{\rm IL}}^{r_{\rm OL}}\,|k|\,dr$, with $|k|=|D|^{1/2}/c$) 
is rather small. In the presence of
wave absorption at the corotation resonance (when $\nu\neq 0$),
we find (assuming $\Theta_{\rm IIa}\gg 1$ and $|\nu|\ll 1$),
\be
|{\cal R}|^2-1\simeq \exp(-2\Theta_{\rm II})+2\pi\nu\exp(-2\Theta_{\rm IIa}),
\ee
and super-reflection can be much more prominent. From Fig.~5 we see that 
the transmission coefficient is generally much smaller than ${\cal D}_c$.
Thus the reflectivity depends mainly on the wave damping at corotation.

Equation (\ref{eq:scc}) or (\ref{eq:S}) clearly 
shows that for $\nu>0$, the wave damping term due to corotation resonance
is positive, ${\cal D}_c > 0$. 
This can be understood from the fact that for $\nu>0$ the Rossby wave
region lies outside $r_c$ (see Fig.~\ref{fig3}), and the dissipation
at the corotation singularity carries away positive energy (just like the
transmitted wave) so that energy conservation [see
eq.~(\ref{eq:Econs})] requires $-1 = -|{\cal R}|^2 + |{\cal T}|^2 +
|{\cal D}_c|$. In this case the corotation singularity enhances
super-reflection, as seen in Figs.~4-5.
On the other hand, for $\nu < 0$, the Rossby wave zone lies inside
$r_c$ (Fig.~\ref{fig2}) and the dissipation carries away negative
energy (like the reflected wave) so that $-1 = -|{\cal R}|^2 + |{\cal
T}|^2 - |{\cal D}_c|$. In this case the corotation singularity tends
to reduce super-reflection, and there is a competition between the
effect of the corotation amplifier [the first term in eq.~(\ref{eq:R})]
and the effect of the corotational absorption [the second term]. To
obtain $|{\cal R}| > 1$ we require
\be
\nu > -{1 \over 2\pi} {1 - {1 \over 4}e^{-2\Theta_{\rm II}} \over
e^{2\Theta_{\rm IIb}} - {1\over 4} e^{-2\Theta_{\rm IIb}}} \simeq -{1
\over 2\pi} e^{-2\Theta_{\rm IIb}},
\label{eq:ineq}\ee
where the second inequality applies in the limit of $\Theta_{\rm IIb}
\gg 1$. This puts a constraint on the disc thickness and the specific 
vorticity slope (note that the sound speed $c$ enters into both $\nu$ and
$\Theta_{\rm IIb}$) in order to achieve super-reflection. 
For example, for a given $c$, the inequality (\ref{eq:ineq}) determines
the critical value of $p$ for which $|{\cal R}|=1$.

Figures 4-5 also reveal an intriguing oscillatory behavior of the reflection,
transmission and damping coefficients.
This non-monotonic behavior may be qualitatively 
understood from the Rossby wave zone 
around $r_c$ (see Figs.~2-3). The WKB wavenumber $k_r$ in the near
vicinity of $r_c$ is given by
\be
k_r^2\simeq -{\kappa^2\over c^2}+{2m\Omega\over r\tomega}{d\over dr}\ln
\zeta={\kappa^2\over c^2}\left(-1+{2\beta\nu r\over r-r_c}\right).
\ee
For $\nu>0$, the Rossby wave zone lies between $r_c$ and $r_c+\Delta r_R$,
where $\Delta r_R=2\beta\nu r_c$.
For quasi-normal modes to be ``trapped in" the Rossby wave zone they must obey the 
the Sommerfeld ``quantization'' condition\footnote{The phase factor $\pi/2$
arises from a detailed analysis of the wave behavior at $r=r_c$
and at $r=r_c+\Delta r_R$: The former gives a phase of $-3\pi/4$ and the latter
gives $\pi/4$.}
$\int_{r_c}^{r_c+\Delta r_R}\!\!dr\, k_r=\pi \nu\sim n\pi+{\pi/2}$,
where $n=0,1,2,\cdots$. Thus when $\nu\simeq n+1/2$, the wave propagating in 
the Rossby zone (Fig.~3) 
is maximally reflected back to the singularity, leading to
maximum negative damping and enhanced net reflection,
as seen in Fig.~\ref{fig5}.
For $\nu<0$ the Rossby wave zone lies between $r_c-|\Delta r_R|$
and $r_c$ (Fig.~2), the wave in the Rossby zone is mostly 
absorbed at the corotation singularity. The asymmetry in the
$\nu>0$ case and the $\nu<0$ case can also be seen in the toy problem of
resonant tunneling (Appendix B). Note that for thin Keplerian disks, $|\nu|$ is
much less than unity for reasonable density profiles ($|p|\sim 1$),
so this non-monotonic behavior is of no practical interest.

\subsection{Numerical Calculation of Reflectivity and Improved WKB Treatment}

\begin{figure}
\centering
\includegraphics[width=13cm]{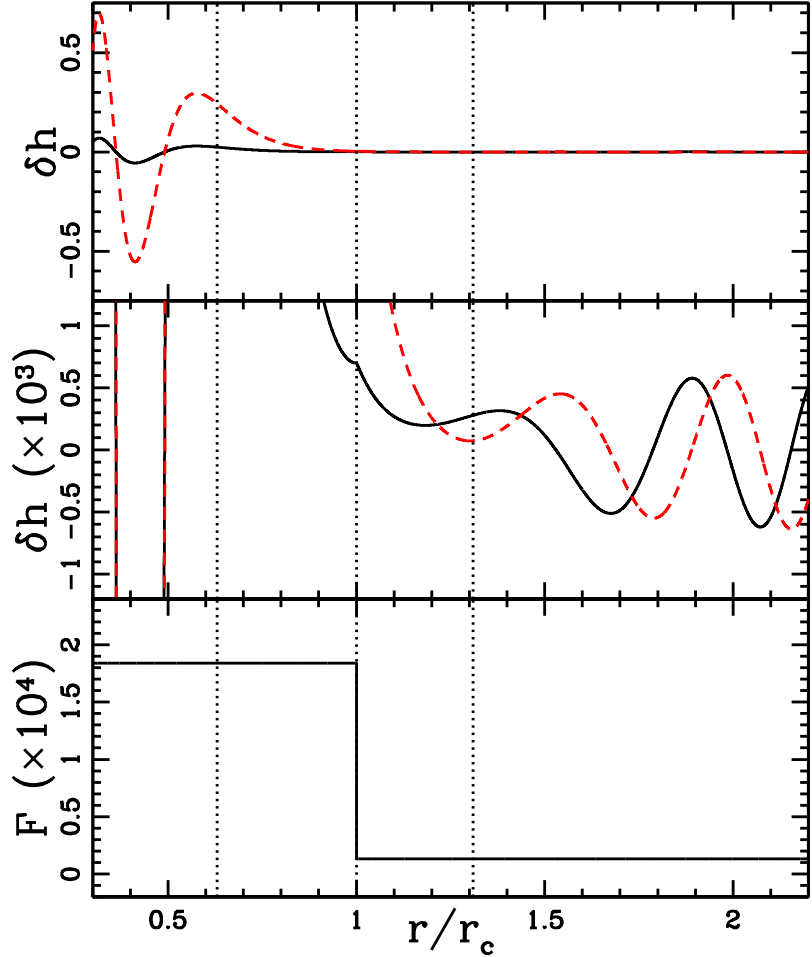}
\caption{Numerical calculation of wave reflection and transmission across corotation for a Keplerian disc with $\nu = 0.033$ and $\beta = 0.1$. The top and middle panel show the enthalpy perturbation $\delta h$, with the solid lines depicting the real components and the dashed lines the imaginary components. Note that there is a discontinuity in the derivative of $\delta h$ at the corotation resonance, resulting in an absorption of flux at the corotation. The bottom panel shows the angular momentum flux carried by the wave [see eq. (67)]. Note that $F(r)$ is conserved away from corotation and the discontinuity in $F$ at $r=r_c$ results from wave absorption. The Lindblad and corotation resonances are indicated by three vertical dotted lines.}
\label{fig6}
\end{figure}

Our analytical expressions derived in section 4-5
are based on global WKB analysis and involves several approximations.
In particular, in our treatment of the corotation resonance
(section 3.2), we assumed $c\ll r\Omega$ --- if this is not satisfied,
some of the neglected terms must be included in equation (\ref{pnearcr})
and the numerical values of our solution may be modified. Thus our results
for $\beta\go 0.1-0.2$ should be treated with caution. 

To assess the validity of our WKB analysis, we also carry out 
computation of the reflectivity by numerical integration of
equations (8)-(10). The outgoing wave boundary condition, 
equation (\ref{eq:outside}),
is imposed at some radius $r_{\rm out} \gg r_{\rm OL}$, such that
\be
\delta h' = \left(ik+{S'\over 2S}-{k'\over 2k}\right)\delta h,
\ee
where $'$ specifies derivatives with respect to $r$.
At some inner radius $r_{\rm in}\ll r_{\rm IL}$, the solution takes the form of
equation (71). The reflection coefficient can be obtained from
\be
|{\cal R}| = \left|\frac{\left[\left(\frac{S'}{2S} -  \frac{k'}{2k} \right) - i k\right]\delta h - \delta h'}
{\left[\left(\frac{S'}{2S} -  \frac{k'}{2k} \right) + i k\right]\delta h - \delta h'}\right|_{r_{\rm in}}
\ee
while the transmission is
\be
|{\cal T}| = \left|\frac{2k}
{\left[\left(\frac{S'}{2S} -  \frac{k'}{2k} \right) + i k\right]\delta h - \delta h'}\right|_{r_{\rm in}} |\delta h|_{r_{\rm out}}~.
\ee
Because of the singularity at corotation, we include a small
positive $\omega_i={\rm Im}(\omega)$ for the frequency so that singularity can be
avoided. Figure \ref{fig6} depicts an example calculation for a Keplerian disc with $\beta = 0.1$ and $p = 2$ (so that $\nu = 0.033$).

\begin{figure}
\centering
\includegraphics[width=13cm]{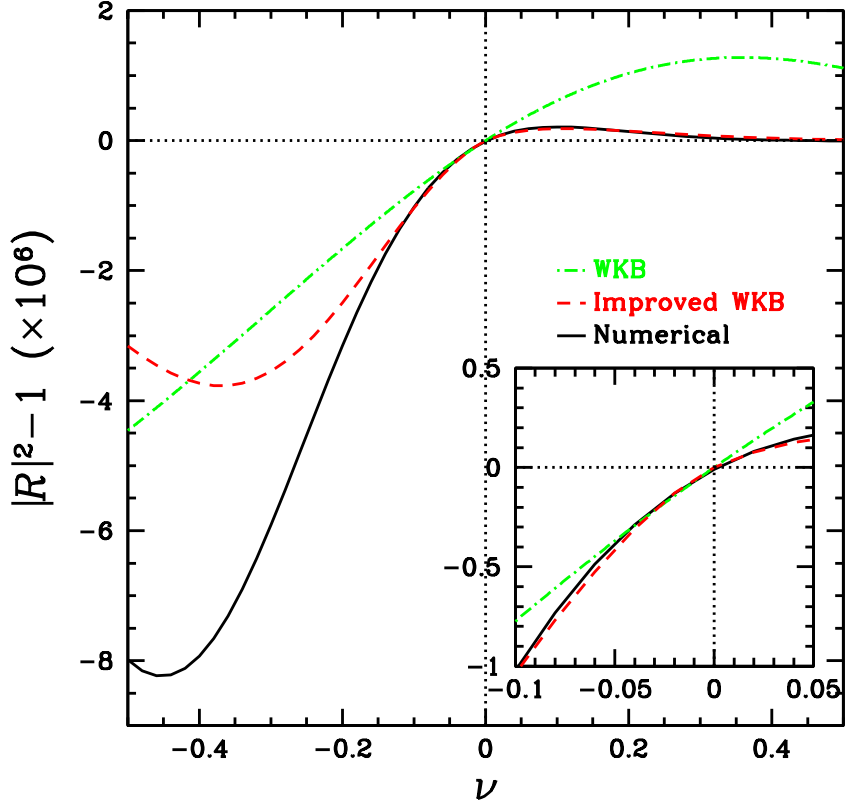}
\vskip -1cm
\caption{The reflection coefficient as a function of $\nu$ calculated using the WKB, improved WKB, and numerical methods for a scale free Keplerian disk with sound speed $c = 0.05 \Omega r$.}
\label{fig7}
\end{figure}

Figure \ref{fig7} shows our numerical result compared with the calculation
from the WKB analysis. There is qualitative agreement between these
results. In particular, both the numerical and WKB results show that wave
absorption at corotation plays the dominant role in determining $|{\cal R}|$,
and wave transmission is unimportant even for small (but nonzero) $\nu$.

However, the WKB solution matches the numerical result closely only
for $|\nu| \ll 1$, indicating that our WKB analysis can be improved.
For the Keplerian disk considered here, the variation in $\nu$ was
achieved by changing the background density index $p$. The WKB results
shown in Figs.~\ref{fig5}-\ref{fig6} assume that the quantities
$\Theta_{\rm IIa}$ and $\Theta_{\rm IIb}$ are not effected by changing
$\nu$. However $p$ plays a non-negligible role in determining the
effective wave number away from the corotation and Lindblad
resonances. Indeed, in obtaining equation \eqref{LRdiffeq} or
\eqref{pnearcr} from \eqref{perturbeq2}, we have neglected several
terms that are negligible near $r_c$ or $r_{\rm IL/OL}$, but
nevertheless important away from these resonances.  Noting that our connection 
formulae [equations (32)-(35) and (42)-(43) involve the asymptotic expansions of 
local solutions around the resonances, 
we can improve our WKB results by adopting the following ansatz: we modify the 
integrands in $\Theta_{\rm IIa}$ and $\Theta_{\rm IIb}$ to include these dropped
terms, 
\ba 
\Theta &=& \int \sqrt{-k_{\rm eff}^2} dr,\\ k_{\rm eff}^2
&=& -\frac{D}{c^2} - \frac{m^2}{r^2} - \frac{p^2 - 1}{4r^2} -
D^{1/2}\left(\frac{d}{dr}D^{-1/2}\right)\frac{1-p}{2}.  
\ea 
Note that
in the expression above, we have left out the singular term ($\propto
\tomega^{-1}$) at corotation and the dominant double singular term
($\propto D^{-2}$) at the Linblad resonances since they have already
been accounted for by the connection formulae in sections 3. As shown
in Fig. \ref{fig6}, the improved WKB result matches the numerical
solution for a much larger range of $\nu$.
The increased values of
$\Theta_{\rm IIa}$ and $\Theta_{\rm IIb}$ for larger $|p|$ act to
suppress super-reflection, and drive reflection coefficient 
toward $|{\cal R}| = 1$.

\section{Global Overstable Modes}

As mentioned in Section 1, global disc instabilities are often related to 
the super-reflection at the corotation barrier. To illustrate this,
we consider a simple boundary condition 
\be
\delta h(r_{{\rm in}}) = 0 
\ee
at the inner radius of the disc, $r_{\rm in} \ll r_{\rm IL}$. The outgoing 
boundary condition at $r>r_{\rm OL}$ implies that the solution 
in the region $r<r_{\rm IL}$ is given by 
\be
\delta h = \sqrt{S/k}\exp\left(-i\int_{r_{\rm IL}}^{r} k
dr+\pi/4\right)+ {\cal R}\sqrt{S/k}
\exp\left(i\int_{r_{\rm IL}}^{r}kdr-\pi/4\right).\label{eq:rinside} 
\ee
Letting ${\cal R} = |{\cal R}| e^{i\varphi}$, and applying the boundary condition (\ref{eq:rinside}) yields the eigenvalue condition:
\be
\tan\left(\int_{r_{\rm in}}^{r_{\rm IL}} k dr 
- \pi/4-\varphi/2\right) = -i\left(\frac{|{\cal R}|-1}{|{\cal R}|+1}\right)~.\label{eq:eigcond}
\ee
Noting that for complex eigenvalue $\omega = \omega_r + i \omega_i$, the wavenumber $k$ is also complex
\be
k = k_r + ik_i = \frac{1}{c} \sqrt{(\tomega_r + i\tomega_i)^2 - \kappa^2} \simeq \frac{1}{c} \sqrt{\tomega_r^2 - \kappa^2} + i \frac{\omega_i\tomega_r}{c\sqrt{\tomega_r^2 -\kappa^2}},
\ee
where we have assumed $|\omega_i| \ll |\omega_r|$. $\tomega_r = \omega_r - m\Omega$ ($< 0$ for region I) and thus the real part of the eigenvalue condition gives
\be
\int_{r_{\rm in}}^{r_{\rm IL}} {1 \over c} \sqrt{(\omega_r - m\Omega)^2 - \kappa^2} dr - \pi/4 -\varphi/2= n\pi~,
\ee
where $n$ is an integer.
The imaginary part of  (\ref{eq:eigcond}) gives (assuming $|k_i| \ll |k_r|$)
\be
\int_{r_{\rm in}}^{r_{\rm IL}} k_i dr \simeq  -\left(\frac{|{\cal R}|-1}{|{\cal R}|+1}\right)~,
\ee
from which we find the growth rate
\be
\omega_i = \left(\frac{|{\cal R}| -1}{|{\cal R}|+1}\right)\left[\int_{r_{\rm in}}^{r_{\rm IL}} |\tomega_r|/c\sqrt{\tomega_r^2 - \kappa^2} dr\right]^{-1}.
\ee
Thus the modes are overstable ($\omega_i > 0$) for $|{\cal R}| > 1$, and stable ($\omega_i < 0$) for $|{\cal R}| < 1$. 

\section{Conclusion}

In this paper we have derived 
explicit expressions for the reflection coefficient, transmission
coefficient and wave absorption coefficient 
when a wave is scattered by the corotation barrier in a disc.
These expressions include both the effects of corotation amplifier 
(which exists regardless of the gradient of the vortensity 
$\zeta=\kappa^2/\Omega\Sigma$ of the
background flow) and wave absorption at the corotation (which depends on 
$d\zeta/dr$). They demonstrate clearly that the corotation wave absorption plays a
dominant role in determining the reflectivity and that the sign of
$d\zeta/dr$ determines whether the corotation singularity enhances or 
diminishes the super-reflectivity. 
Our result can be understood in terms of the location of the Rossby wave zone 
relative to the corotation radius. We also carried out numerical calculations
of the reflectivity. Our result provides the conditions 
(in terms of disc thickness, rotation
profile and surface density profile) for which super-reflection is 
achieved and global overstable modes in discs are possible.

In future works we will explore global oscillation modes and their
stabilities in a variety of astrophysical contexts, ranging from
accreting white dwarfs to accreting black hole systems. The possible 
overstabilities of these modes are directly
linked to the effects studied in this paper and may provide 
explanations for some of the 
quasi-periodic variabilities observed in these systems.

\section*{Acknowledgments}

DL thanks Peter Goldreich and Ramesh Narayan for useful conversations.
This work has been supported in part by NASA Grant NNX07AG81G, NSF
grants AST 0707628, and by {\it Chandra} grant TM6-7004X
(Smithsonian Astrophysical Observatory). 

\appendix

\renewcommand{\theequation}{A-\arabic{equation}}
  \setcounter{equation}{0}  

\section{Stokes Phenomenon and the Matching Condition Across the Corotation
Singularity}

As we saw in section 3.1 the perturbation equation near the
corotation resonance can be solved in terms of the Whittaker
function. However in order to provide the matching conditions we must
carefully consider the effect of the Stokes Phenomenon on the
asymptotic expansions.

Stokes phenomenon causes the functional form of the asymptotic expansion of an entire
function to be different at different points in the complex plane.
The general asymptotic solution of the Whittaker equation 
for $|z|\gg 1$ can be written as a linear combination of two functions
\be
P(z)=e^{z/2}z^{-\nu},\qquad 
Q(z)=e^{-z/2}z^\nu,
\ee
i.e., $AP(z)+BQ(z)$. 
However, because of the Stokes phenomenon, the coefficients $A$ and $B$ 
can change when crossing the Stokes lines. For Whittaker functions,
the Stokes lines are the positive real axis (where $P$ is dominant and
$Q$ is sub-dominant) and negative real axis (where $Q$ is dominant and
$P$ is sub-dominant).

Consider specific solution to the Whittaker equation, with the asymptotic
expansion ($|z|\gg 1$) given by
\be
F(z)\rightarrow AP(z)+BQ(z),\qquad ({\rm for}~~\arg(z)=0)
\label{eq:f}\ee
on the real axis. Our goal is to derive the expansion coefficients 
of $F(z)$ on the negative real axis ($\arg(z)=\pi$). 
To achieve this, we use the general results obtained by Heading (1962): 

(1) ``The coefficient of the subdominant term after crossing the Stokes
    line $=$ the coefficient of the subdominant term before crossing
    the Stokes line + $ T_n \times$ the coefficient of the dominant
    term on the Stokes line.''

(2) ``The coefficient of the subdominant term on the Stokes line $=$ the
    coefficient of the subdominant term before the Stokes line $+
    {1\over 2} T_n \times$ the coefficient of the dominant term on the
    Stokes line.''

Here $T_n$ is the Stokes multiplier for crossing the Stokes line $\arg(z) = n\pi$ in the direction of increasing $\arg(z)$ given by
\footnote{Note that a typo in equation (18) of Heading (1962) has 
been corrected here.}
\be
 T_n ={2\pi i e^{-2\pi ins\nu}\over \Gamma(s\nu)\Gamma(1 + s\nu)}
\ee
where $s = (-1)^n$. Note that since the pole of equation (\ref{Whittakerdiffeq}) lies below the real axis, $\arg(z)$ increases as we we move along the contour from $z$ positive and real ($\arg(z) = 0$) to $z$ negative and real ($\arg(z) = \pi$).

Since equation (\ref{eq:f}) is given on one of the Stokes line,
we first determine $F(z)$ in the region $-\pi<\arg(z)<0$:
\be
F(z)\sim  AP(z)+\left(B-{1\over 2}  T_0 A\right)Q(z),
\qquad ({\rm for}~ -\pi<\arg(z)<0).
\ee
Then in region $0<\arg(z)<\pi$ we have
\be
F(z)\sim  AP(z)+\left(B+{1\over 2} T_0 A\right)Q(z),
\qquad ({\rm for}~ 0<\arg(z)<\pi).
\ee
Thus on the negative real axis, we obtain
\be
F(z)\sim 
\left[A+{1\over 2} T_1
\left(B+{1\over 2}  T_0 A\right)\right]P(z)+\left(B+{1\over 2} {\rm T_0}A\right)Q(z),
\qquad ({\rm for}~ \arg(z)=\pi).
\ee

The connection formulae for the two independent Whittaker functions for $|z| \gg 1$ are
\be
\psi_- = W_{\nu,1/2}(z)\sim
\Biggl\{
\begin{array}{ll}
Q(z) & ~~~\arg(z)=0\\
Q(z)+{1\over 2}{\rm T_1} P(z) & ~~~\arg(z)=\pi
\end{array}
\ee
\be
\psi_+ = e^{-i\pi\nu}\textrm{W}_{-\nu, 1/2} (ze^{-i\pi}) + \frac{1}{2} T_0 \textrm{W}_{\nu,1/2} (z). 
\sim
\Biggl\{
\begin{array}{ll}
P(z) & ~~~\arg(z)=0\\
{1\over 2}T_0 Q(z)+
\left(1+{1\over 4} T_0 T_1\right)P(z) & ~~~\arg(z)=\pi,
\end{array}
\ee
where $P(z)$ and $Q(z)$ have the asymptotic behavior
\ba
P(z) = e^{+z/2 -\nu \log(z)} &\sim&
\Biggl\{\begin{array}{ll}
\exp\left(\int_{r_c}^{r}\! \tilde k \,dr \right) 
& \qquad \qquad \qquad ~~\textrm{for } r \gg r_{c}\\
e^{-i\pi\nu}\exp\left(-\int_r^{r_{c}}\! \tilde k \,dr \right) 
&\qquad \qquad \qquad~~\textrm{for } r\ll r_{c}~.
\end{array}\\
Q(z) = e^{-z/2 + \nu \log(z)} 
&\sim&
\Biggl\{\begin{array}{ll}
\exp\left(-\int_{r_c}^{r}\! \tilde k \,dr \right) 
& \qquad \qquad \qquad \qquad~\textrm{for } r \gg r_{c}\\
e^{i\pi\nu}\exp\left(\int_r^{r_{c}}\! \tilde k \,dr \right) 
&\qquad \qquad \qquad\qquad~\textrm{for } r\ll r_{c}~.
\end{array}
\ea
for $|z| \gg 1$, since in this limit $|z| \gg \log |z|$.

\section{Resonance Tunneling}

\renewcommand{\theequation}{B-\arabic{equation}}
  \setcounter{equation}{0}  

To help understand the effect of wave absorption at the corotation
singularity we consider the following toy problem. For a 1-D
quantum mechanical potential of the form
\be
V(x) = {C\over x},
\ee
we have the time-independent wave equation
\be
\left(\frac{1}{2}\frac{d^2}{dx^2} + E - \frac{C}{x + i\epsilon}\right)
\psi = 0,
\label{eq:wave}\ee
where we have set the Planck constant and particle mass to unity.
In equation (\ref{eq:wave}), a small imaginary part $i\epsilon$ is added 
to x, to account for the physical requirement of a growing incoming 
perturbation. The sign of $\epsilon$ must be the same as the sign of $C$, 
so as to give us the correct physical behavior.
For concreteness we will assume $C>0$ (and thus $\epsilon > 0$), 
though the same calculation can be performed for $C < 0$ ($\epsilon < 0$) 
with similar results. 

Let $y = 2ik(x + i\epsilon)$, where $k = \sqrt{2E}$, we have
\be
\frac{d^2\psi}{dy^2} + \left(-\frac{1}{4} + \frac{\nu}{y} \right)\psi = 0,
\qquad {\rm with}~~\nu=iC/k, 
\label{qmwhittaker},
\ee
which we recognize as the Whittaker differential equation.
With $C > 0$, $y$ lies in the domain $\pi/2 \leq \arg(y) \leq 3\pi/2$.
Using the same procedure as discussed in Appendix A, we obtain the 
connection formulae for the general solution of 
equation \eqref{qmwhittaker} as
\be
\psi \rightarrow  \Biggl\{\begin{array}{ll} 
A P(z) + B Q(z)
&\qquad \textrm{for } \arg(z) = \pi/2 \\
(A + T_1 B) P(z) + B Q(z)
&\qquad \textrm{for } \arg(z) = 3\pi/2.
\end{array}
\ee

Consider a wave propagating from $x=-\infty$ and impinging on the
potential. The transmitted wave at $x>0$ has the form
\be
\psi_+ \rightarrow  e^{y/2} y^{-\nu} = e^{ikx -i(C/k)\ln|2kx| + \pi C/2 k}
\qquad \textrm{for } x \gg 1/k\\.
\ee
The corresponding wave solution in the $x<0$ region has the asymptotic form
\be
\psi_+ \rightarrow e^{y/2} y^{-\nu} = e^{ikx -i(C/k)\ln|2kx| + 3\pi C/2k}
\qquad \textrm{for } x \ll -1/k.
\ee
This gives 
\be
|{\cal R}| = 0,\qquad |{\cal T}| = e^{-\pi C/ k}
\ee
for a wave incident from $x < 0$.

Now consider a wave incident from the $x > 0$ region towards small $x$.
The transmitted wave is given by 
\be
\psi_- \rightarrow e^{-y/2} y^{+\nu} = e^{-ikx +i(C/k)\ln|2kx| - 3\pi
C/2k}\qquad \textrm{for } x \ll -1/k.
\ee
Connecting to the $x>0$ region, we have
\be
\psi_- \rightarrow e^{-y/2} y^{+\nu} - T_1 e^{y/2} y^{-\nu}= e^{-ikx
+i(C/k)\ln|2kx| - \pi C/2k} - T_1 e^{ikx -i(C/k)\ln|2kx| + \pi C/2k}
\qquad \textrm{for } x \gg 1/k.
\ee
This gives the reflection  and transmission coefficients
\ba
&& |{\cal R}| =e^{\pi C/k}|T_1|=
 \frac{2\pi e^{-\pi C/ k}}{|\Gamma(-iC/k) \Gamma(1 -
iC/k)|} 
= 2 e^{-\pi C/k} \sinh(\pi C/k) = 1- e^{-2\pi C/k}, \\
&&|{\cal T}| = e^{-\pi C/k},
\ea
for a wave incident from the positive side of the singularity
[a similar result was obtained by Budden (1979) in the context of
wave propagation in cold plasma].

We can define the wave absorption coefficient at the singularity as
\be
{\cal D} = 1- |{\cal T}|^2 - |{\cal R}|^2
\ee
For the forward moving incident wave ($+$) and the backwards moving incident
wave ($-$), we have
\be
{\cal D}_+ = 1 - e^{-2\pi C/k},  \qquad {\cal D}_- = e^{-2\pi C/k} \left(1 - e^{-2\pi C/k} \right).
\ee



\begin{thebibliography}{99}

\bibitem[]{} Abramowitz, M, Stegun, I.A. 1964, Handbook of
Mathematical Functions (Dover: New York)

\bibitem[]{}
Balbus, S.A., Hawley, J.F. 1998, Rev. Mod. Phys., 70, 1.

\bibitem[]{}
Budden, K.G., Phil. Trans. R. Soc. Lond. A, 1979, 290, 405

\bibitem[]{}
Dickenson, R.E., 1968, J. Atmos. Sci., 25, 984

\bibitem[]{}
Goldreich, P. 1988, in ``Origin, structure and evolution of galaxies''
(Proceedings of the Guo Shoujing Summer School of Astrophysics, 
Tunxi, China), ed. L.Z. Fang (World Scientific: Singapore), p.~127

\bibitem[]{}
Goldreich, P., Lynden-Bell, D. 1965, MNRAS, 130, 125

\bibitem[]{}
Goldreich, P., Goodman, J., Narayan, R. 1986, MNRAS, 221, 339

\bibitem[]{}
Goldreich, P., Tremaine, S. 1979, ApJ, 233, 857

\bibitem[]{}
Goodman, J., Evans, N. 1999, MNRAS, 309, 599

\bibitem[]{}
Heading, J. 1962, J. Lond. Math. Soc., 37, 195

\bibitem[]{}
Julian, W.H., Toomre, A. 1966, ApJ, 146, 810

\bibitem[]{}
Kato, S., 2003, PASJ, 55, 257

\bibitem[]{}
Li, H., Finn, J.M., Lovelace, R.V.E., Colgate, S.A. 1999, ApJ, 533, 1023

\bibitem[]{}
Li, L., Goodman, J., Narayan, R., 2003, ApJ, 593, 980

\bibitem[]{}
Lin, C.C., Lau, Y.Y. 1975, SIAM J. Appl. Math., 29, 352

\bibitem[]{}
Lovelace, R.V.E., Li, H., Colgate, S.A., Nelson, A.F. 1999, ApJ, 513, 805

\bibitem[]{}
Lynden-Bell, D., Kalnajs, A.J. 1972, MNRAS, 157, 1

\bibitem[]{}
Mark, J.W.K. 1976, ApJ, 205, 363

\bibitem[]{}
Narayan, R., Goldreich, P., Goodman, J. 1987, MNRAS, 228, 1

\bibitem[]{}
Papaloizou, J.C.B., Lin, D.N.C. 1995, ARAA, 33 505

\bibitem[]{}
Papaloizou, J.C.B., Pringle, J.E. 1984, MNRAS, 208, 721

\bibitem[]{}
Papaloizou, J.C.B., Pringle, J.E. 1987, MNRAS, 225, 267

\bibitem[]{}
Pedlosky, J. 1987, Geophysical Fluid Dynamics. Springer-Verlag, Berlin

\bibitem[]{}
Shu, F.H. 1992, The Physics of Astrophysics II: Gas Dynamics. University
Science Books, Mill Valley, CA, chapter 12

\bibitem[]{}
Shu, F.H., Laughlin, G., Lizano, S., Galli, D., 2000, ApJ, 535, 190

\bibitem[]{}
Tagger, M., Pellat, R. 1999, A\&A, 349, 1003

\bibitem[]{}
Tagger, M., 2001 A\&A, 380, 750

\bibitem[]{}
Tagger, M., Varniere, P. 2006, ApJ, 652, 1457

\bibitem[]{}
Tanaka, H., Takeuchi, T., Ward, W.R. 2002, ApJ, 565, 1257

\bibitem[]{}
Zhang, H., Lai, D., MNRAS, 2006, 368, 917

\end{thebibliography}
\end{document}